# The MICADO first light imager for the ELT: Testing the Lyot coronograph prototypes

Pierre Baudoz*[a], Elsa Huby[a], Olivier Dupuis[a], Faouzi Boussaha[b], Yann Clénet[a], Ric Davies[c]

[a] LESIA, Observatoire de Paris, Université PSL, CNRS, Université Paris Cité, Sorbonne Université, 5 place Jules Janssen, 92195 Meudon, France
[b] GEPI, Observatoire de Paris, Université PSL, CNRS, 77 Avenue Denfert-Rochereau, Paris, 75014, France
[c] Max-Planck-Institut für extraterrestrische Physik, Garching, Germany

## ABSTRACT

MICADO, the European Extremely Large Telescope first light imager will feature a dedicated high contrast imaging mode specifically designed for observing and characterizing exoplanets and circumstellar disks. Its improved sensitivity and angular resolution, compared to existing instruments will significantly increase our knowledge on these planetary systems. MICADO will include three classical Lyot coronagraphs, one vector phase-apodized pupil plane (vAPP) and two sparse apertures. After rapidly describing the final design of MICADO high contrast mode, we will describe the current state of development of the coronographic components and the testing of the first Lyot coronagraph prototypes.

## 1. INTRODUCTION

Instruments equipped with extreme adaptive optics, such as SPHERE on the VLT [1], GPI on GEMINI [2] or SCEXAO on Subaru[3], have greatly enhanced our understanding of planetary system formation and evolution. However, many questions remain unanswered, especially on the formation scenarios of these planetary systems and the frequency of of giant planets in distant orbits beyond 5-10 AU. Space instruments aboard the JWST provide new insights into exoplanetary science [4] and the coronagraph planned on the Roman Space telescope should broaden the bandwidth coverage by the end of this decade [5]. But current ground-based and future space instruments remain limited in angular resolution compared to instruments designed for the ELT. The development of a specialized instrument for direct imaging and spectroscopy of exoplanets on the ELT, such as ESO's Planetary Camera and Spectrograph (PCS), will only begin after the initial instruments are operational. Exploiting the fivefold improvement in angular resolution of the ELT over current ground-based or planned space-based telescopes offers a valuable opportunity for exoplanetary research before the expected deployment of dedicated ELT planet finders in the mid to late 2030s.

Because of its large central obscuration and its segmented pupil, the ELT pupil is not optimized for planet direct observation using coronagraphy. Indeed, the large central obscuration and the presence of gaps increase the number of edges in the pupil, thus the amount of stellar light diffracted by these edges. One solution to overcome this problem is to couple the focal plane mask with an apodization mask at the entrance pupil of the instrument to compensate for the effect of the diffraction by the central obscuration. However, the final design of MICADO does not permit it. Ideed, unlike SPHERE for example, MICADO is not exclusively focused on exoplanet science. Therefore, MICADO's optical design represents a balance between its various operational modes.

*pierre.baudoz@obspm.fr

## 2. MICADO HIGH CONTRAST MODE

In this context, simple Classical Lyot Coronagraphs (CLC), without apodization are selected as a baseline for MICADO. A CLC is an occulting mask in the focal plane and must be coupled with a Lyot stop located in a pupil plane downstream of the focal plane mask.

To take full advantage of the high angular resolution of the ELT and allow different observation conditions, we decided to select three CLC that allow the detection of planet at short distance from the star (coronagraph with a small Inner Working Angle, IWA). The selection of the size of the focal plane masks is linked to the selection of the dimension of the Lyot stop. The sizes of the 2 elements (radius of the focal mask and filtering of the Lyot stop) are optimized at the same time for the three components. The final design includes two CLC with a radius of the focal plane masks corresponding to 2 $\lambda/D$ at the wavelength of 1.3 µm (J band) and 2.2 µm (K band) and a third CLC with a radius of 4 $\lambda/D$ at 2.2 µm (K band). For more information on this optimization, see Perrot et al. 2019 [6] and Baudoz et al. 2023 [7]. We also studied the improvement when adding a more complex coronagraph like a vortex coronagraph [7] to allow reaching even smaller angular separation to the star. We found that such coronagraphs are too limited by the large central obscuration of the ELT and the fact that, in MICADO, the atmospheric dispersion compensator (ADC) is located after the entrance focal plane where the focal plane masks take place.

To mitigate this issue, one solution is to use pupil plane components like phase or amplitude apodization that are less sensitive to jitter and can be located after the ADC, thus are not limited by atmospheric dispersion. A pupil phase apodization uses interferences to attenuate the diffraction wings in a limited area of the field of view without decreasing the total transmission of the instrument. This apodization decreases the Strehl ratio (SR) and an optimization needs to be done to maximize the SR and the useful FOV while minimizing the diffracted light at the shortest angular separation. In MICADO, we choose a vector Apodizing Phase Plate (vAPP, [8]) with a rectangular shape, an inner working angle of 2.6 $\lambda/D$, an outer working angle of 20 $\lambda/D$ and a theoretical contrast of $10^{-4}$ at 3 $\lambda/D$. A first vAPP prototype has been fabricated for visible wavelengths by the company ColorLink Japan, Ltd [9]. MICADO will also include two sparse aperture masks (SAM), which insert in the pupil an opaque mask with a number of holes spatially distributed so that each pair of holes will contribute to a unique spatial frequency. The optimization of the two SAM designs, which include either 9 or 18 holes, is described in Huby et al. 2024 [10]. The MICADO high contrast mode are summarized in Table 1 and the location of each component inside the MICADO cryostat is indicated in Figure 1. More information on the high contrast mode of MICADO, the estimated performance and the simulation tool to estimate the expected images can be found elsewhere [7][11][12].

| Coronagraph | IWA or equivalent | Spectral bands | Position in the cryostat | Notes |
| --- | --- | --- | --- | --- |
| CLC15 | 15 mas | J to K | Entrance focal wheel. Used with Lyot stops | Limited zenith angle for large spectral bands |
| CLC25 | 25 mas | J to K | Entrance focal wheel | Limited zenith angle for large spectral bands |
| CLC50 | 50 mas | J to K | Entrance focal wheel | |
| vAPP | 2.6 $\lambda/D$ | Narrow bands J to K | Exit pupil wheel | OWA = 20 $\lambda/D$ |
| SAM9 | 0.5 $\lambda/D$ | Narrow bands J to K | Exit pupil wheel | Sparse aperture masking, 9 holes |
| SAM18 | 0.5 $\lambda/D$ | Narrow bands J to K | Exit pupil wheel | Sparse aperture masking, 18 holes |

Table 1: Summary of the different elements of the high-contrast imaging mode of MICADO

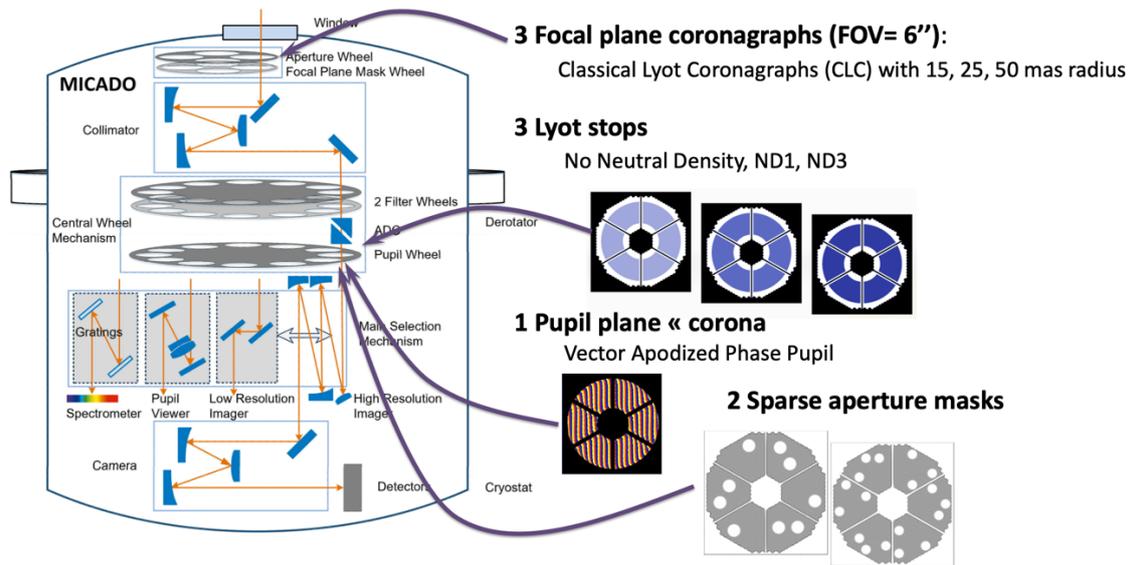

Figure 1. MICADO cryostat showing the position of the different components of the high contrast imaging mode

## 3. CLASSICAL LYOT CORONAGRAPH DESIGN

This paper is mainly dedicated to the description of the three Classical Lyot Coronagraph (CLC) focal plane masks that are planned to take place in the entrance focal plane wheel of MICADO. Their rough characteristics and description can be found below.
The occulting spots of the CLC coronagraphs are planned to be opaque masks, deposited on a transparent substrate (IR fused silica). There are three CLC masks, one (CLC15) optimized for smaller inner working angle (15 mas) at short wavelength (J,H), one (CLC25) optimized for intermediate inner working angle (25 mas) and one optimized (CLC50) for deep search detection at large working angle (>50mas) and for observation at large airmass and with degraded AO correction (faint target for example). The only difference between the 3 components is the occulting spot diameter that is deposited at the centre of the substrate (Figure 2).

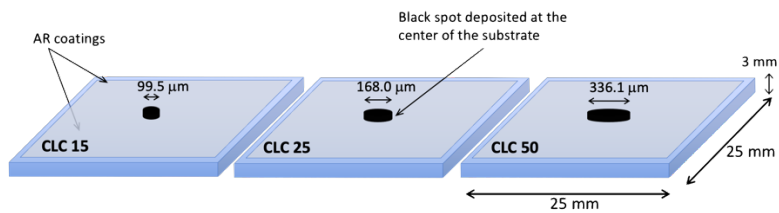

Figure 2. Comparison between the substrates of CLC15, CLC25, CLC50

The optical specifications for the CLC are given in Table 2. For the occulting spot transmission, we tested 2 different materials (TiN and Cr) and decided for TiN since the needed optical density requires only a thin layer of 300 nm of TiN (see Figure 7) compared to a layer of more than 500 nm for Cr. To hold the optical substrate between warm temperature (up to 323 K) and cryostat temperature (down to 80K), we developed a mounting assembling made of the same aluminum than the entire focal plane wheel and using metal flex blades to allow the thermal expansion difference between the mount and IR fused silicate substrate (see Figure 3).

| Parameter | Value |
|---|---|
| **Substrate size** | 25 mm x 25 mm ±0.01mm |
| **Substrate thickness** | 3 mm ±0.05mm |
| **Substrate parallelism** | < 1' |
| **Substrate material** | IR Silicate |
| **Substrate transmission quality** | < 15 nm RMS over the clear aperture |
| **Scratch/dig** | 10/5 |
| **Optical clear aperture** | >20 mm x 20 mm (6"x6") |
| **Anti-Reflection on both sides** | <0.5% from 1.15 to 2.3µm |
| **Occulting spot diameter** | 99.5 µm, 168 µm or 336.1 µm |
| **Occ. spot diameter tolerance** | ± 2 µm |
| **Occ. spot centring tolerance** | +/- 0.1 mm |
| **Occulting spot optical density** | >4 |
| **Operation temperature** | 80 K to 120 K |

Table 2. Focal plane mask optical specifications

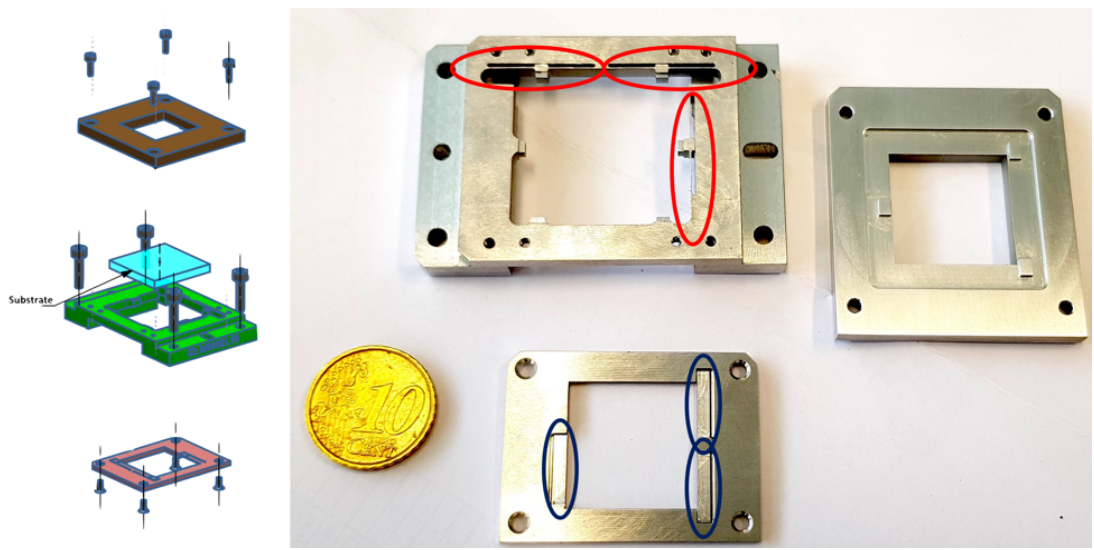

Figure 3. Left: Assembly plan of the mount. Right: Picture of the 3 parts of a prototype mount. Flex blades used to hold the components are shown in red for the 3 lateral ones and in dark blue for the 2 top ones.

## 4.  CLC DEVELOPMENT TESTS

To fabricate and validate the CLCL components, the plan is to test both mechanical and optical behaviors using mechanical and optical prototypes before the fabrication of the final elements. The different steps described in Table 3 and Table 4 are based on a mechanical prototype on one hand (shown in Figure 3) to test the feasibility of the mechanical mount (especially the flex blades) and thermal constraint effect on the mount and on the optical substrate. On the other hands, optical prototypes of the CLC have been built to verify the quality of the deposition of the central occulting TiN spot and compare the measured performance with the simulated one.

To test the coronagraphic performance of the optical components, we have developed an IR bench that simulates the optical interface of the CLC on MICADO with an ELT pupil, an entrance F/D of 17.75, the expected Lyot stop filtering and a series of IR filters (J, H, K band) placed in front of an IR camera as shown in Figure 4.

| Task | Mechanical Validations | Date |
|---|---|---|
| **M1** | Prototype mount fabrication | ✓ Validated |
| **M2** | Prototype thermal cycling | ✓ Validated |
| **M3** | Cold opto-mechanical measurements | 10/2024 |
| **M4** | Final mounts thermal cycling | 03/2025 |

Table 3. Mechanical validation plan

| Task | Optical Validations | Date |
|---|---|---|
| **O1** | Spot quality (absorption, diameter) | ✓ Validated |
| **O2** | Performance of optical prototypes | ✓ Partly validated |
| **O3** | Mounted optical quality of final substrates | ✓ Validated |
| **O4** | Optical quality at cold T° (M3 above) | 10/2024 |
| **O5** | Overall performance of final components | 12/2024 |

Table 4. Optical validation plan

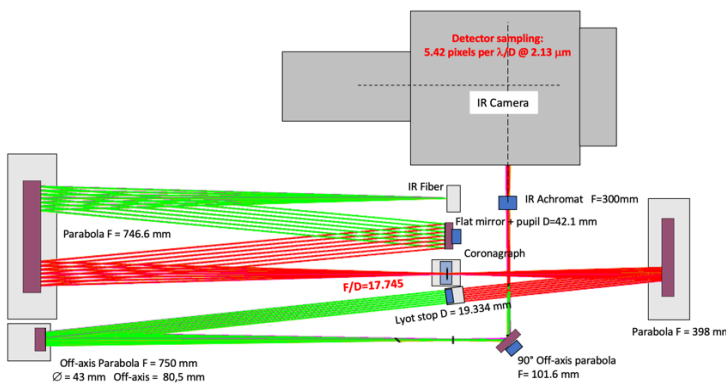
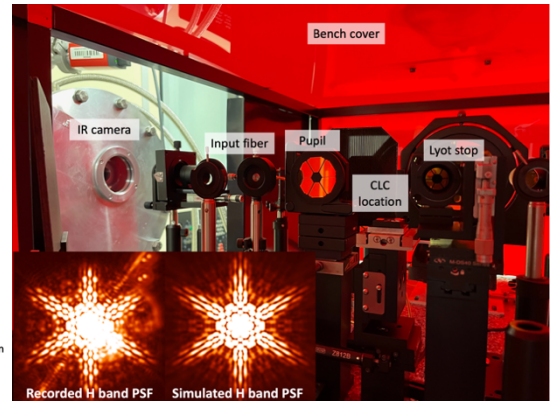

Figure 4. Left: Optical design of the CLC validation bench. Right: Picture of the bench showing the main components. Insert on the right: Comparison of a recorded PSF with the simulated one.

**Mechanical validation of the CLC prototype mount**

The task M1 and M2 have been validated using the prototype shown in Figure 3. To verify that the flex blades do not undergo non reversible deformation, we mount and dismount the optical substrate at room temperature (warm tests) and measure the distance of the blade with respect to the external frame using an optical microscope (Figure 5). This distance is the same before and after the substrate is mounted. We then redo the test at cold temperature applying temperature variation to the mount with a substrate mounted inside it. The thermal cycling applies a warm temperature at 323K and reaches 4 times the cold temperature of 80K for more than 2h for each of them as shown in Figure 6. The distance between the blade and the external frame is shown in Figure 5 for both warm and cold tests. The distance without mounted substrate is recovered after each test with a precision of a few micrometers which is the precision of the method. Thus, the flex blades do not undergo non-reversible deformation even at cold temperature when the difference of expansion between fused silicate and aluminum is the largest.

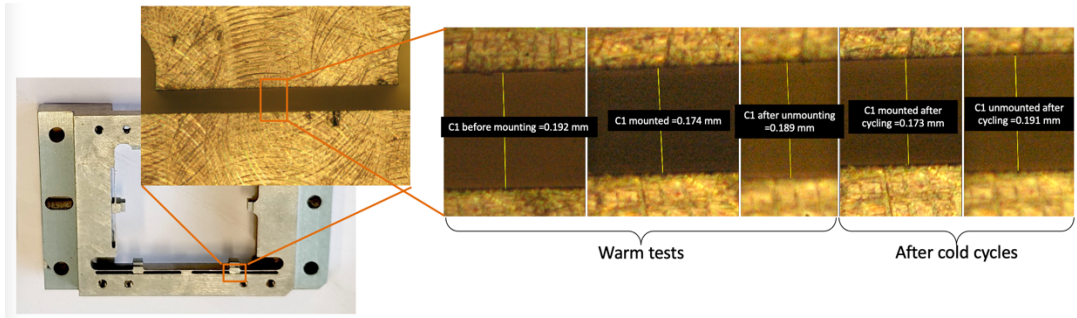

Figure 5. Measurements of the prototype mount before and after thermal cycling showing that the flex blades are retrieving their initial place after cold cycling (Tasks M1 & M2). Images are recorded using an optical microscope. Scale in the images is calibrated using an image of precise metallic ruler.

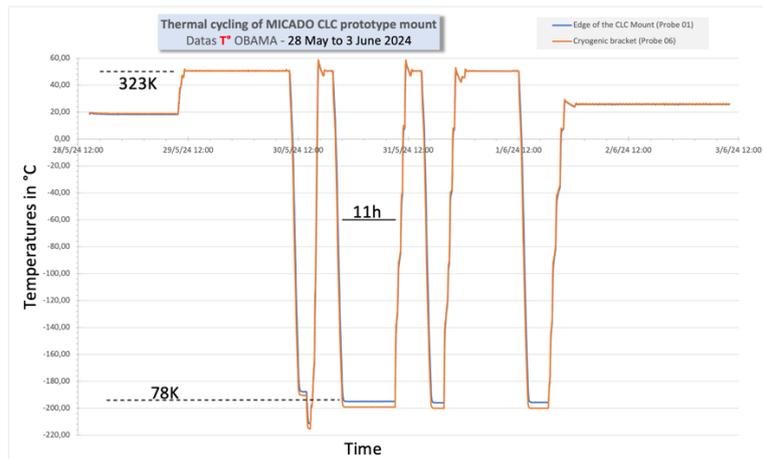

Figure 6. Cold cycling timeline showing 4 cooling time lasting more than 2h and reaching below 80K.

**Optical validation of the prototype mount and CLCs**

The validation of the quality of the occulting spot (Task O1) is done measuring the size of the spot using micrometric measurements under an optical microscope (Figure 7, Left). The thickness of the TiN deposited is measured during the deposition procedure. The thickness that needs to be deposited to reach the required optical density of 4 (OD=4, $10^{-4}$) is estimated previously on separate components using the same deposition recipe on a complete substrate to avoid the diffraction effect of the limited spot size (Figure 7, Middle). The curves show that 300 nm of TiN is required to reach OD=4 for the three wavelength bands.

The optical quality of the substrates when mounted at warm temperature (Task O3) is tested by recording interferometric measurements of the transmitting substrate mounted and unmounted. Both measurements show the same value of 12 nm RMS over the full 20 mm x 20mm clear aperture (Figure 7, Right). Thus, the mount does not create aberration on the substrate at warm temperature and the substrate is within the required optical quality specifications.

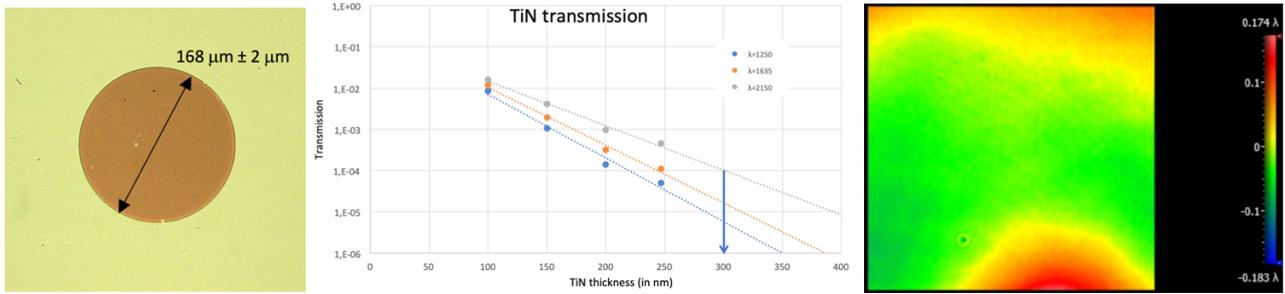

Figure 7. Left: Measurements of a CLC25 spot (O1). Middle: Transmission of TiN as a function of its deposited thickness for 3 wavelengths corresponding to the J, H and K band. (Validation of *Task O1*). Right: aberrations of a mounted component (12 nm RMS) over a surface of 20mm x 20 mm, O3.

**Performance validation of the prototype CLCs**

We started the performance validation (O2) on the testbed by first measuring the transmission of a point source as a function of the distance to the optical axis. As shown in Figure 8, the measured values are fitting well the theoretical curves. CLC15 is slightly off as shown in Figure 8 and Figure 9 in J band but the focalization of the component was not done properly. New tests will be carried out after a proper focalization to verify if this effect can explain these degraded results.

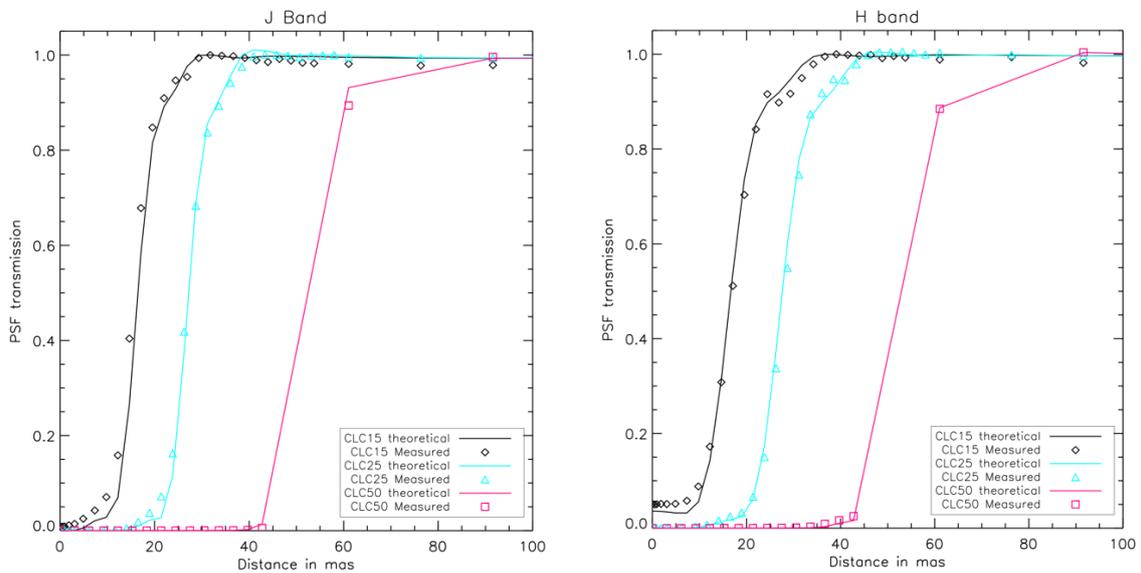

Figure 8. Throughput as a function of the distance to the axis in mas (O2 validation).

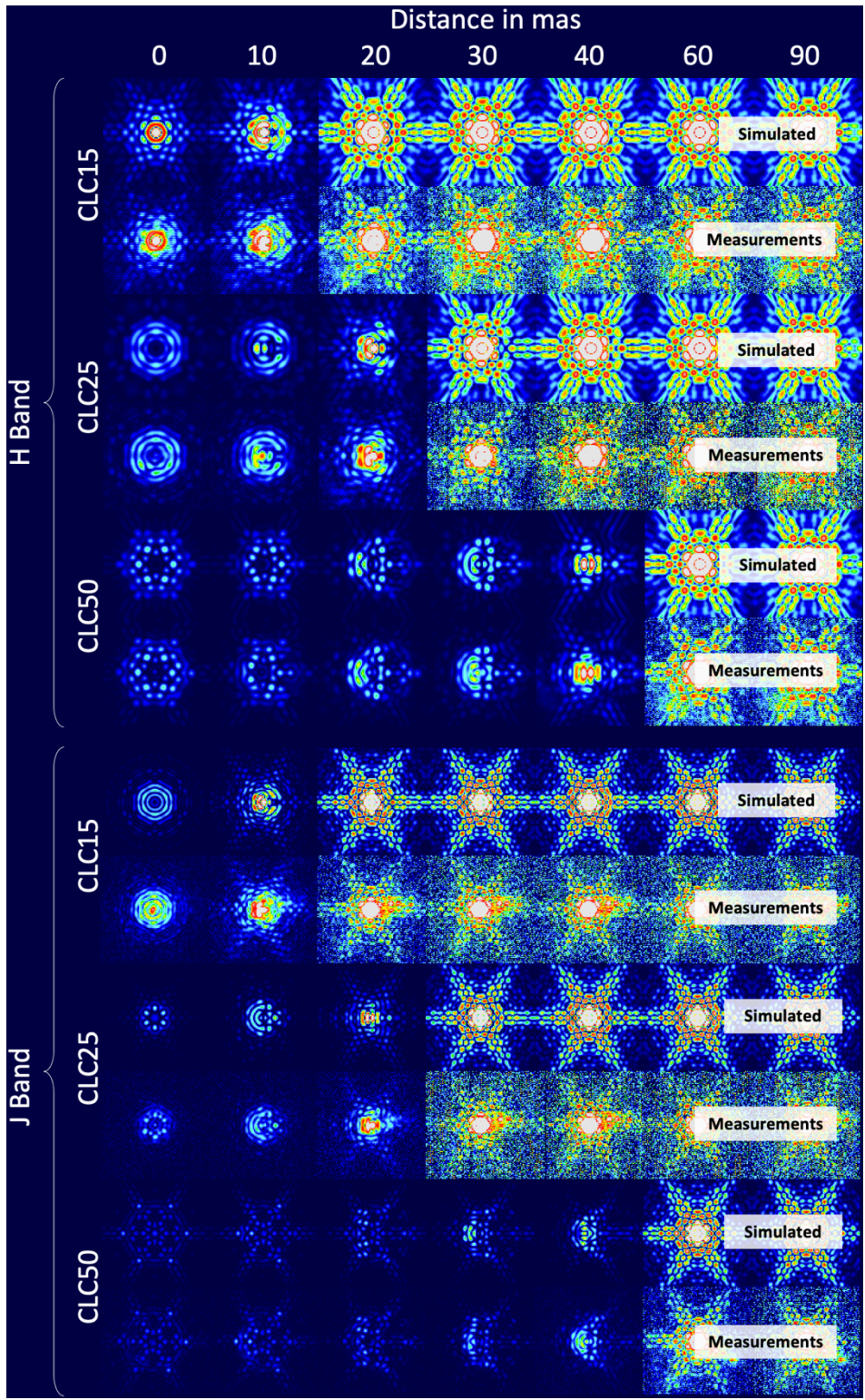

Figure 9. Typical images of the 3 CLCs for several typical distances. Measured and simulated images are compared for each coronagraphs at J and H bands.

## 5. CONCLUSION

In this paper, we showed that the development of the three classical Lyot coronagraph (CLC) planned for MICADO, the first-light imaging instrument of the ELT, is well on track. We built a mechanical prototype mount and CLC prototypes. The mechanical mount is behaving as expected at warm and cold temperature. To test the performance of the CLC in terms of coronagraphy, we developed a dedicated IR testbed to simulate the ELT pupil and F/D entrance. The first measured performance is close to the numerical simulated performance. More tests will be carried out in the next months with the measurements of the aberrations possibly introduced by the mount at cold temperature (Tasks M3/O4) and more tests of the CLC prototypes, especially CLC15. After the validation of these tests, we will move on to the fabrication of the final mounts and optical elements before doing the final components verifications planned in March 2025.

## ACKNOWLEDGEMENTS

This work has benefited from the support of the French Programme d'Investissement d'Avenir through the project F-CELT ANR-21-ESRE-0008 and by the Action Spécifique Haute Résolution Angulaire (ASHRA) of CNRS/INSU co-funded by CNES.

## REFERENCES

[1] Beuzit, J.-L. et al., "SPHERE: the exoplanet imager for the Very Large Telescope," Astronomy and Astrophysics 631 (2019). doi:10.1051/0004-6361/201935251

[2] Macintosh, B. et al., "The Gemini Planet Imager: looking back over five years and forward to the future.", Proceedings of Adaptive Optics for Extremely Large Telescopes (AO4ELT6), 10703 (2018). doi:10.1117/12.2314253

[3] Currie, T. et al., "Direct imaging and astrometric detection of a gas giant planet orbiting an accelerating star," Science 380, 198–203. (2023) doi:10.1126/science.abo6192

[4] Boccaletti, A. et al., "Imaging detection of the inner dust belt and the four exoplanets in the HR 8799 system with JWST's MIRI coronagraph," Astronomy and Astrophysics, 686 2024). doi:10.1051/0004-6361/202347912

[5] Bailey, V. P et al., "Nancy Grace Roman Space Telescope coronagraph instrument overview and status," Proceedings of SPIE Conference 2023 - 12680, 126800T (Oct. 2023).

[6] Perrot, C., Baudoz, P., Boccaletti, A., Rousset, G., Huby, E., Clénet, Y., Durand, S., Davies, R., "Design study and first performance simulation of the ELT/MICADO focal plane coronagraphs," Proceedings of Adaptive Optics for Extremely Large Telescopes (AO4ELT5), (2019), arXiv e-prints. doi:10.48550/arXiv.1804.01371

[7] Baudoz, P., Huby, E., Vidal, F., Gendron, E., Clenet, Y. and Davies, R., "Design and performance of the MICADO high contrat mode," Proceedings of Adaptive Optics for Extremely Large Telescopes (AO4ELT7), 092 (2023). doi:10.13009/AO4ELT7-2023-092

[8] Doelman, D. et al., "Vector-apodizing phase plate coronagraph: design, current performance, and future development," Applied Optics 60, D52 (2032). doi:10.1364/AO.422155

[9] Doelman et al., ''vector-Apodizing Phase Plates: prototyping for ELT/METIS/MICADO", Proceedings of SPIE Conference 2024 – Poster 13100-257 (2024)

[10] Huby et al., ''The MICADO first light imager for the ELT: Sparse Aperture Masks, design and simulations," Proceedings of SPIE Conference Series 2024 –13096-203 (2024)

[11] Baudoz, P., Huby, E., Clénet, Y., "Exoplanetary systems study with MICADO", Proc. of the Annual meeting of the French Society of Astronomy and Astrophysics (SF2A), (2019)

[12] Huby et al., "The MICADO first light imager for the ELT: MISTHIC simulation pipeline for the high contrast mode of MICADO", Proceedings of SPIE Conference Series 2024–13097-225 (2024)